\documentclass[preprint2]{aastex}

\newcommand\vect[1]{\mbox{\boldmath{$ #1 $}}}

\shortauthors{Yahagi \& Yoshii}
\shorttitle{N-body code with AMR}
\begin{document}
\title{$N$-Body Code with Adaptive Mesh Refinement}
\author{Hideki Yahagi\altaffilmark{1} and Yuzuru Yoshii\altaffilmark{2}}
\affil{Institute of Astronomy, University of Tokyo, Tokyo 181-0015, Japan}
\email{hyahagi@ioa.s.u-tokyo.ac.jp, yoshii@ioa.s.u-tokyo.ac.jp}
\altaffiltext{1}{Research Fellow of the Japan Society for the Promotion of 
Science}
\altaffiltext{2}{Also at the Research Center for the Early Universe, 
University of Tokyo, Tokyo, 113-0033, Japan}
\begin{abstract}
We have developed a simulation code with the techniques which enhance
both spatial and time resolution of the PM method for which the
spatial resolution is restricted by the spacing of structured
mesh. The adaptive mesh refinement (AMR) technique subdivides the
cells which satisfy the refinement criterion recursively. The
hierarchical meshes are maintained by the special data structure and
are modified in accordance with the change of particle distribution.
In general, as the resolution of the simulation increases, its time step
 must be shortened and more computational time is required to complete
 the simulation. Since the AMR enhances the spatial resolution locally,
 we reduce the time step locally also, instead of shortening it
 globally. For this purpose we used a technique of hierarchical time
 steps (HTS) which changes the time step, from particle to particle,
 depending on the size of the cell in which particles reside.
Some test calculations show that our implementation of AMR and HTS is
 successful. We have performed cosmological
simulation runs based on our code and found that many of halo objects
have density profiles which are well fitted to the universal profile
proposed by Navarro, Frenk, \& White (1996) over the entire range of
their radius.
  
\end{abstract}
\keywords
{
	galaxies: formation ---
	large-scale structure of universe ---
	methods: n-body simulations ---
	methods: numerical
}
%##################################################
% Section 1: Introduction
%##################################################
\section{Introduction \label{sec:intro}}
%Estimation of required spatial dynamic range
Simulating the formation process of galaxies in clusters and fields
simultaneously is one of the promising strategies to study how
statistical properties of observed galaxies have originated. For this
purpose, the box size in which the simulation is performed must be
larger than a typical cluster separation, which is a few tens to a
hundred Mpc. At the same time, in order to reproduce the intrinsic
dynamical structure of each simulated galaxy, the Newtonian gravity
should be calculated on a scale well below the characteristic length
of individual galaxies, which is a few kpcs. Hence, for the simulation
of formation of multiple galaxies, the ratio of box size to minimum
resolved scale or the spatial dynamic range must be greater than $10^5$.

%Estimation of particle number
In addition, there is another dynamic range in $N$-body systems often
called the mass dynamic range, or simply, the number of particles if
the mass of particles is the same. Strictly speaking, dark matter
component evolves according to the collisionless Boltzmann
equation. However, since computer resources are finite, the
distribution function of ideal collisionless component is replaced by
a finite sum of localized distribution functions corresponding to the
finite number of particles in the $N$-body simulation. This causes
two-body relaxation which artificially affects the evolution of dark
matter distribution. The effect of relaxation can be suppressed by
introducing more particles. We crudely estimate how many particles are
required to trace the evolution of galactic dark halos in a cosmological
simulation by comparing the relaxation time scale of dark halos and
their age. Our naive estimation gives $N\sim10^9$ at the minimum (see
Appendix \ref{sec:mass_res}).

%AMR is efficient
It is instructive to examine whether the particle-mesh (PM) method is
capable of the required spatial dynamic range of $10^5$ and the required
mass dynamic range of $10^9$. The spatial dynamic range of the PM method is
restricted by the mesh spacing. When the number of meshes is taken as
many as the number of particles, the spatial dynamic range is about
$10^3$ for a simulation with $10^9$ particles, which is two orders
smaller than required. However, a simulation with $10^{15}$ meshes and
particles, whose spatial dynamic range is about $10^5$, is not only
unrealistic today, but also inefficient. The adaptive mesh refinement
(AMR) is one of the prescriptions to resolve this problem by
introducing finer meshes hierarchically in regions where higher
spatial resolution is required (Berger \& Oliger 1984).

%Other adaptive PM codes
First, Villumsen (1989) introduced cubic subgrids hierarchically into the PM 
method to increase its spatial resolution. While his hierarchical PM code 
places adaptive meshes by hand, subgrids are automatically located in the 
particle--multiple mesh code developed by Jessop, Duncan, \& Chau (1994).
Gelato, Chernoff, \& Wasserman (1997) extended their methods to handle
isolated $N$-body systems. Their code is applicable to non-cosmological
problems. On the other hand, adopting the multigrid method as the Poisson
solver, Suisalu \& Saar (1995) developed a code which can treat rectangular
adaptive meshes. Furthermore, Kravtsov, Klypin, \& Khokhlov (1997) developed
the adaptive refinement tree (ART) code which subdivides all cells which 
satisfy the user defined refinement criterion regardless of the shape of 
the boundary between coarse meshes and refined meshes.

Calculating the force between particles is the most time consuming
part in $N$-body methods. In order to overcome this problem, Aarseth (1963)
changed the timing to calculate the force, from particle to particle,
depending on the time scale of force variation. This technique, named
individual time steps  is implemented in the ART code (Kravtsov et
al. 1998) with modification such that the time steps of particles are
determined by the local density, though not described in detail in their
paper. We also adopted the modified individual time steps, or
hierarchical time steps (HTS) so that our code is adaptive not only
in space but also in time, or our code is four dimensionally
adaptive. Furthermore, our code is designed to incorporate the
hydrodynamics code with AMR (Yahagi \& Yoshii 2000), similar to the
codes developed by Anninos, Norman, \& Clarke (1994) and Norman \& Bryan 
(1999). While the refined regions are restricted to be rectangular
parallelepipeds in their codes, this limitation has been relaxed in
ours.

%Outline of the paper
In \S \ref{sec:code} we review the basic equations for the cosmological 
$N$-body simulation and describe our code in detail putting emphasis on the 
parts related to AMR and HTS. Details of tests and the results are 
discussed in \S \ref{sec:test}. Finally, we summarize this work in \S 
\ref{sec:summary}.

%##################################################
% Section 2: Description of the code
%##################################################
\section{Description of the code\label{sec:code}}
% Section 2.1: Basic equations
\subsection{Basic equations}
For the cosmological simulation, the comoving coordinate system is suitable 
to represent particle position:
\begin{eqnarray}
	\vect x = a^{-1} \vect r,
\end{eqnarray}
where $a$ is the scale factor, and $\vect x$ and $\vect r$ are comoving and 
proper positions, respectively. However, in order to incorporate the 
hydrodynamics code, the proper coordinate system is suitable to represent 
the velocity:
\begin{eqnarray}
	\vect u = \frac{d\vect r}{dt} - \frac{\dot a}{a} \vect r.
\end{eqnarray}
Using these variables, the basic equations for $N$-body systems in the 
expanding universe are given by:
\begin{eqnarray}
\frac{d\vect x}{dt} &=& \frac{1}{a} \vect u, \label{eq:velocity}\\
\frac{d (a \vect u)}{dt} &=& \vect g(\vect x), \label{eq:eq_of_motion}\\
\vect g(\vect x) &=& - \nabla \phi(\vect x), \label{eq:force_diff}\\
\triangle \phi(\vect x) &=& \frac{4 \pi G}{a} (\rho(\vect x) - \bar{\rho}), 
\label{eq:poisson}
\end{eqnarray}
and
\begin{eqnarray} 
\dot a &=& H_0 \sqrt{(1-\Omega_0-\lambda_0) + a^{-1} \Omega_0 + a^2 
\lambda_0}. \label{eq:expansion}
\end{eqnarray}
Note that $\rho$ denotes the comoving density, here.
How to discretize these equations depends on the problem to be solved and 
the strategy for it. Since the primary objective of our code is to perform 
the cosmological simulation with a wide dynamic range in both space and 
mass, we must save memory as much as possible. Thus using the variable
time-step leapfrog  integrator which does not require additional
memory, we discretize equations (\ref{eq:velocity}) and
(\ref{eq:eq_of_motion}) as follows:
\begin{eqnarray}
{\vect x}_{i+1} &=& {\vect x}_{i} + \frac{1}{a_{i+1/2}} {\vect u}_{i+1/2} ~ 
\Delta t_i, \label{eq:intg_pos}\\
{\vect u}_{i+1/2} &=& \frac{a_{i-1/2}}{a_{i+1/2}} {\vect u}_{i-1/2} + 
\frac{1}{a_{i+1/2}} {\vect g}_{i} ~ \frac{\Delta t_{i-1} + \Delta t_i}{2}. \label{eq:intg_vel}
\end{eqnarray}

Among a variety of methods to solve equations (\ref{eq:force_diff}) and 
(\ref{eq:poisson}), we adopt the mesh based method, such as the PM method, 
which calculates the force between particles through the following steps:
(i) Mass of particles is assigned to their adjacent grid nodes.
(ii) Discretized Poisson equation is solved to give the potential on each node.
(iii) Force on each node is calculated as the difference of the potential 
on nearby nodes.
(iv) Force on particles is interpolated from those on the adjacent nodes.
We discuss the mass assignment (i), force interpolation (iv), and potential 
differentiation (iii) in \S \ref{sec:interpolation}, while the Poisson 
solver (ii) in \S \ref{sec:poisson}.

%Before going ahead, we describe the integration of the Friedmann equation. A usual method in the literature is to use the Taylor series expansion to the fourth order \citet{he88, anc94}.Instead, we integrate the Friedmann equation by adopting the predictor corrector manner. First, differentiation  of equation (\ref{eq:expansion}) gives the acceleration of the scale factor:
%\begin{eqnarray}
%\ddot{a} = H_0^2 (a \Omega_V - \frac{1}{2} a^{-2} \Omega_{m}). \label{eq:addot}
%\end{eqnarray}
%Then we set predictors as
%\begin{eqnarray}
%\dot{a}_{i+1}^p &=& \dot{a}_{i} + \ddot{a}_{i} \Delta t \label{eq:Frdmn 
%extrp}, \label{eq:Frdmn prdct}
%\end{eqnarray}
%and
%\begin{eqnarray}
%a_{i+1}^p &=& a_{i} + \frac{1}{2}(\dot{a}_{i} + \dot{a}_{i+1}^p) \Delta t.
%\end{eqnarray}
%The correctors are calculated by substituting $\dot{a}_{i+1}^p$ into 
%$\dot{a}_{i+1}$ in equation (\ref{eq:Frdmn extrp}), so that
%\begin{eqnarray}
%a_{i+1}^{c1} &=& a_{i} + \frac{1}{2}(\dot{a}_{i} + \dot{a}_{i+1}^{c1}) 
%\Delta t \label{eq:Frdmn crrct}.
%\end{eqnarray}
%Although the error decreases with iterating the above correction procedure, 
%one iteration suffices for convergence. Moreover, $\Delta t$ in equations 
%(\ref{eq:Frdmn prdct})-(\ref{eq:Frdmn crrct}) need not be the same as 
%$\Delta t$ in equations (\ref{eq:intg_pos}) and (\ref{eq:intg_vel}).

% Section 2.2: Hierarchical mesh
\subsection{Hierarchical mesh\label{sec:hrch mesh}}
\begin{figure}
\plotone{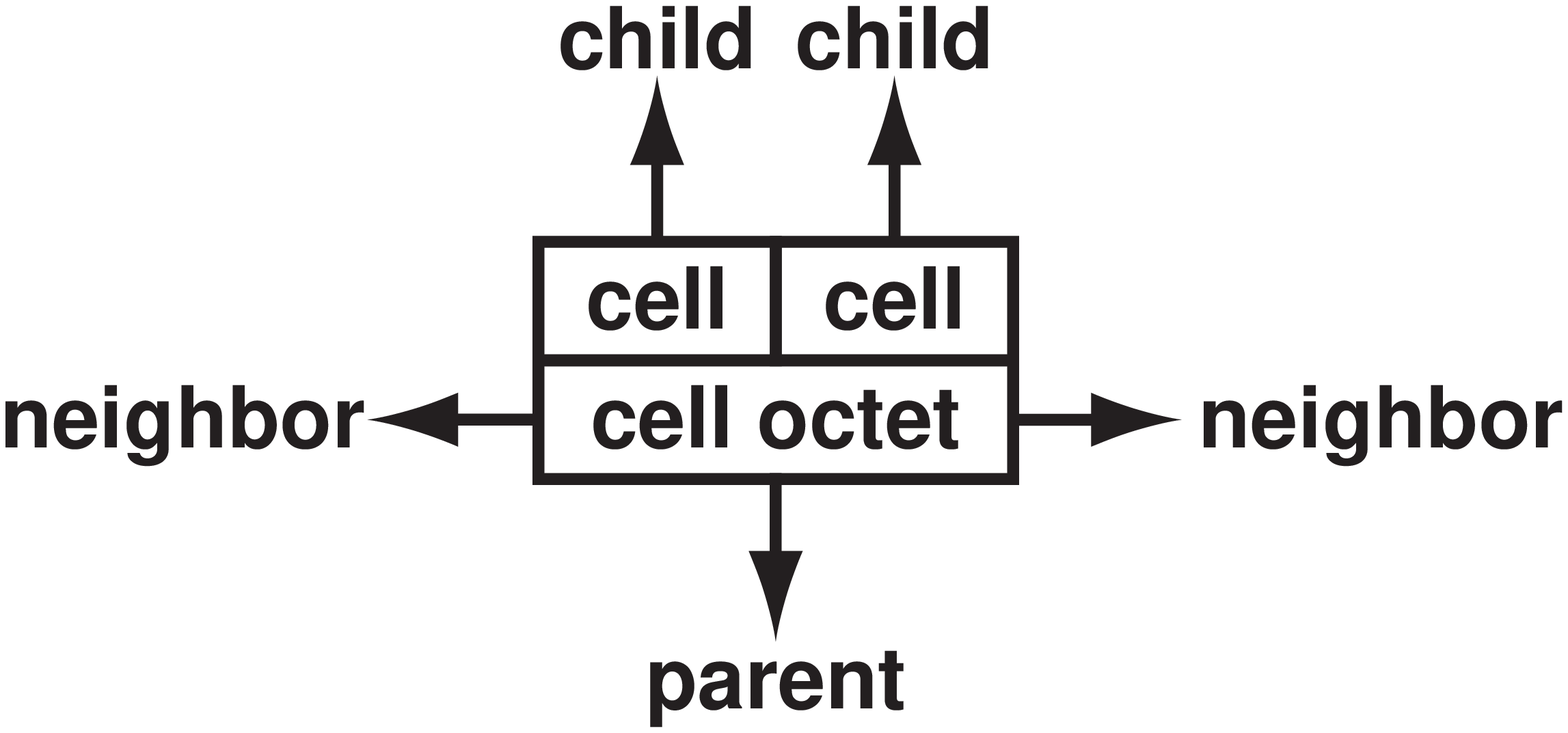}
\caption{
Data structure of the building block which is used to construct the 
hierarchical meshes.  Eight half-sized cells having the same parent are 
grouped into a cell octet and share the pointer to the parent and its six 
neighbors.
\label{fig:data_structure}}
\end{figure}

%Data structure of the PM method.
The PM method uses a cubic mesh to calculate the force on particles. 
Therefore, the spatial resolution of the PM method is limited by the 
spacing of this mesh. Like the ART code developed by \citet{kkk97}, our 
code overcomes this limitation by subdividing the cells in regions where 
higher spatial resolution is required. This adaptive mesh refinement 
corresponds to adding data sets for small cells onto the homogeneous and 
cubic base mesh hierarchically. In order to maintain and modify the 
hierarchical mesh structure, it is required to access the parent, child, 
and neighbor cells. The storage for pointers to the parent and neighbor 
cells can be saved by grouping the cells which have the same parent cell 
\citep{kho98}.
Since one refinement process divides a cell into eight half-sized
cells, we group these eight cells together and call this group a cell
octet (Fig. \ref{fig:data_structure}). Each cell octet keeps a pointer to
the parent cell and pointers to the six neighboring cell octets. In
the $N$-body code, in addition to these data, a pointer to the first
particle in the cell octet, the number of particles in the cell octet,
and the integral position of the cell octet are also stored. 
The integral position is the position rounded off to integer.
While cell octets residing in the same level are connected by the doubly
linked lists in the ART code, our code stores the cell octets having the
same level in an array consecutively. The level of a cell,
$L$, is defined as $L=\log_2(l_0/l_{L})$, where $l_0$ and $l_{L}$ are
the sizes of the simulation box and the cell, respectively. We use
cells having integral level only.
\placefigure{fig:data_structure}

\begin{figure}
\plotone{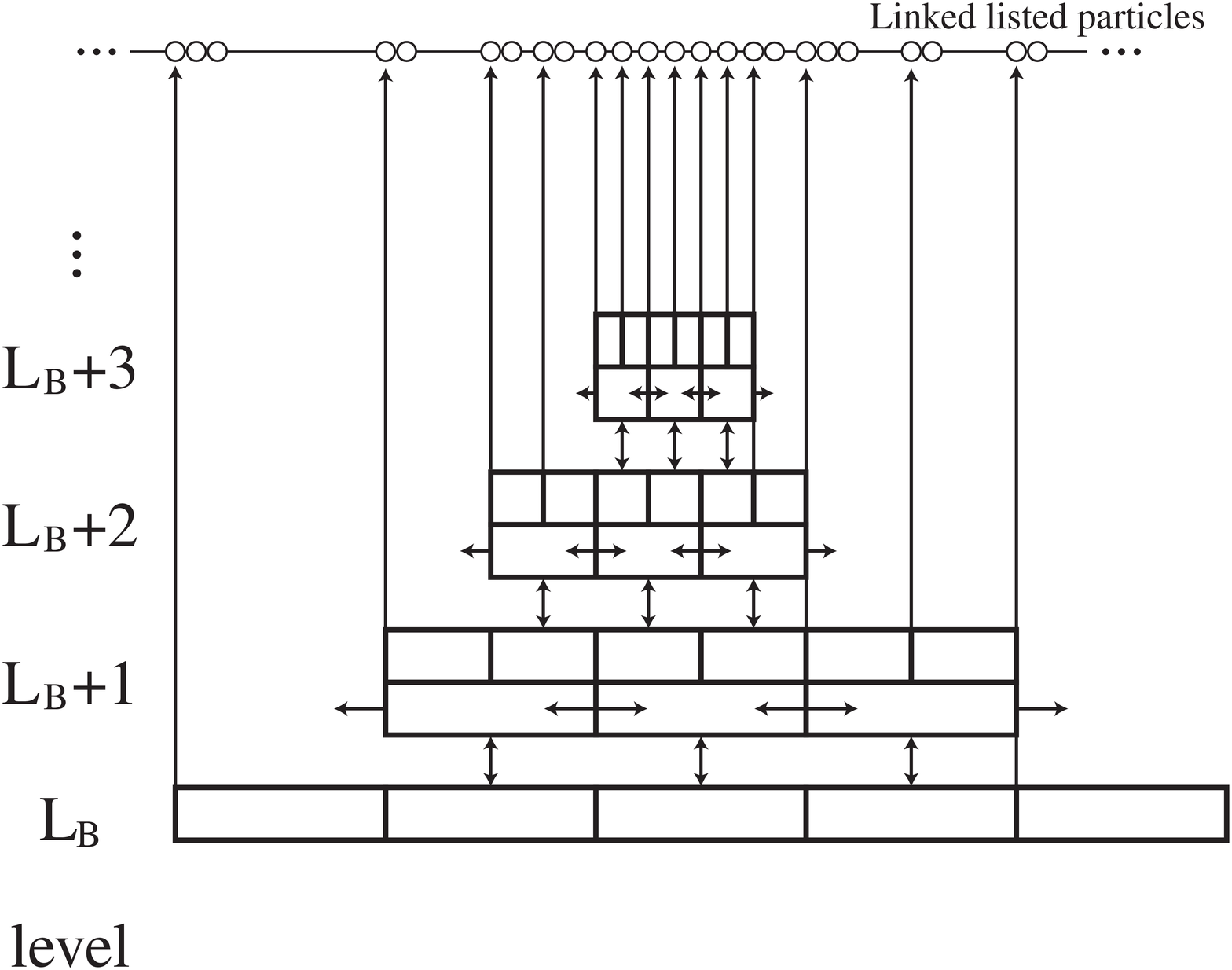}
\caption{
Construction of the hierarchical meshes by connecting cell octets as shown. 
These cell octets are added or removed dynamically as the system evolves.
The cells without the child cell octet use their pointer to the child
octet as the pointer to the head of the linked listed particles which
reside in the cell.
\label{fig:hierarchical_mesh}}
\end{figure}

Moreover, the same level cell octets are sorted by the Morton ordering
(Barnes \& Hut 1989; Warren \& Salmon 1993). The key for the Morton
ordering, $k_M$, is given by
\begin{eqnarray}
k_M &=& \sum_{L_i=0}^{L_D} (2^{3L_i} k^x_{L_i} + 2^{3L_i+1} k^y_{L_i} + 2^{3L_i+2} k^z_{L_i}),
\end{eqnarray}
where $L_D$ is the level of the smallest hierarchical mesh, and
$k^x_{L_i}, k^y_{L_i}$, and $k^z_{L_i}$ are the $L_i$-th bit of the
integral position of the cell octet normalized by $l_{L_D}$
($\tilde{x}, \tilde{y}$, and $\tilde{z}$). Thus,
\begin{eqnarray}
\tilde{x} &=& \sum_{L_i=0}^{L_D} 2^{L_D-L_i} k^x_{L_i},\\
\tilde{y} &=& \sum_{L_i=0}^{L_D} 2^{L_D-L_i} k^y_{L_i},\\
\tilde{z} &=& \sum_{L_i=0}^{L_D} 2^{L_D-L_i} k^z_{L_i}.
\end{eqnarray}
We do not sort particles as Warren and Salmon's hashed oct-tree code
(Warren \& Salmon 1993), because it is incompatible with HTS, though
sorting hierarchical cells is compatible with HTS. The hierarchical
mesh is maintained using the cell octets as the building blocks
(Fig. \ref{fig:hierarchical_mesh}), which are added or removed from
meshes freely.
\placefigure{fig:hierarchical_mesh}

Basically each cell has only one pointer to the child octet, but the
storage changes depending on what physics is incorporated in the
code. In the code for the self gravitating system, density and
potential is stored. No additional storage is required for the
$N$-body code, since the storage for a pointer to the first particle
in the cell is shared by the pointer to the child cell octet. The
hydrodynamics code requires storage for fluid density, specific energy
and fluid velocity in each cell.

\begin{figure}
\plotone{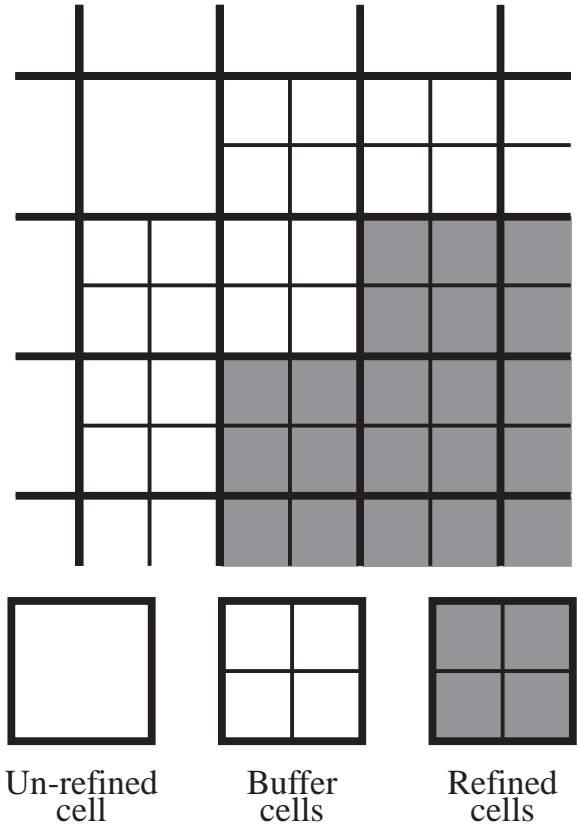}
\caption{
Three types of the hierarchical meshes such as unrefined, refined and 
buffer cells.  All cells which satisfy the refinement criterion are 
subdivided and called refined cells ({\it shaded}). Those cells which are 
next to the refined cells are also subdivided and called buffer cells.  The 
others remain as unrefined cells.
\label{fig:refine_buffer}}
\end{figure}

%Refinement criterion.
The condition for subdividing the cells, or the refinement criterion, plays 
an important role in the AMR. In general context of the AMR, the refinement 
criterion is defined using error estimators such as the residual of the 
partial differential equation to be solved. However, since we introduced 
the AMR to keep the mass dynamic range all the way from homogeneous to 
clustered configurations, we refined all cells which contain 
$N_{\mbox{\small rfn}}$ particles and more as the ART code.
This criterion sets the upper limit to the mass in unrefined cells. We
set $N_{\mbox{\small rfn}}=4$ throughout this paper unless otherwise stated.
%Maximum level of refinement.
Following this refinement criterion, the hierarchical meshes are 
constructed recursively on the cubic structured mesh whose base level is 
$L_B$, until they reach the dynamic range level, $L_D$. At the same
time, the cells which do not satisfy the refinement criterion are removed
from the hierarchical meshes. In addition to the refined cells which
satisfy the refinement criterion, we also subdivide those cells, named
buffer cells, which are adjacent to the refined cells
(Fig. \ref{fig:refine_buffer}). 

\placefigure{fig:refine_buffer}

For the illustrative purpose, we take an example from a slice of the 
cosmological simulation for an LCDM universe described in \S
\ref{sec:lcdm test}. The left panel of Figure \ref{fig:eg mesh} shows
the distribution of particles, and the right panel shows the
configuration of hierarchical meshes placed in accordance with the
above prescription.
While the cells including the sheet structure are refined once, more levels 
of hierarchical meshes are placed in the regions where massive halos reside.
\placefigure{fig:eg mesh}

% Section 2.3: Individual time steps
\subsection{Individual time steps}
%How to determine the time step.
We adopted a time stepping scheme called the hierarchical time steps
(e.g. Makino 1991). The time step of each particle is not the same but
restricted to $2^{-i}$ times of the longest time steps. We chose $i$
to be the refinement level, $L-L_B$, (Kravtsov et al. 1998) so that
the trajectory of the level $L$ particle, which is included in the
level $L$ refined cell but not in any level $L+1$ refined cells, is
integrated by the time interval $\Delta t_{L}$ which is related to the
time interval for the base level $L_B$ particles, $\Delta t_{L_{B}}$,
as
\begin{eqnarray}
\Delta t_{L} = 2^{L_{B}-L} \Delta t_{L_{B}}. \label{eq:indiv_time}
\end{eqnarray}
In our code, $\Delta t_{L_{B}}$ is given by
\begin{eqnarray}
\Delta t'_{L_{B}} &=& \frac{\epsilon l_{L_{B}}}{a^{-1} v_{max}}_{~,}\\ 
\Delta t_{L_{B}}  &=& \frac{\epsilon l_{L_{B}}}{a^{-1} (v_{max} + g_{max} a^{-1} \Delta t'_{L_{B}})}_{~,}  \label{eq:base_time}
\end{eqnarray}
where $\epsilon$ is a free parameter and fixed as $\epsilon = 0.25$ 
throughout this paper.

%mesh structure modification, Poisson solver, velocity step, position step.
Equation (\ref{eq:indiv_time}) indicates that the level $L$ particles step 
twice while the level $L-1$ particles step once. Each step of particles is 
split into three steps such as velocity step by half time interval ($\Delta 
t_L / 2$), position step by full time interval ($\Delta t_L$), and velocity 
step by half time interval ($\Delta t_L / 2$) again.
For example, when the times of hierarchical meshes of level $L_0$ and larger 
synchronize, their mesh structure is modified simultaneously in accordance 
with the refinement criterion. When this synchronization occurs at $t_i$, 
the level $L$ particles, for which $L \geq L_0$, step their velocity
from $t_i - \Delta t_L / 2$ to $t_i + \Delta t_{L'} / 2$, where $L$ and
$L'$ are the particle's levels at $t_i - \Delta t_L$ and $t_i$,
respectively. The force on particles is interpolated from the corners of
the level $L'$ refined cells. Hence, even if the cell to which a particle
belongs is removed, the particle can step their velocity properly. As
far as the particle's levels at $t_i - \Delta t_L$ and $t_i$ are the
same, the integration of its trajectory is the same as that by the
leapfrog method. The pseudo-code describing this flow of procedures is
given in Appendix \ref{sec:pseudo-code}.

% Section 2.4: Mass assignment and force interpolation
\subsection{Mass assignment and force interpolation\label{sec:interpolation}}
In our code the mass of particles is assigned to the
corners of the neighboring cells, and the force on particles is
interpolated from them. Among various assignment and
interpolation schemes for the PM method (see e.g. Hockney \& Eastwood
1988), we adopted the cloud-in-cell (CIC) scheme:
\begin{eqnarray}
\rho_{l,m,n} &=& \sum_{\it all~particles} m_p l_{L_{B}}^{-3} W_{L_{B}}
(\vect x - \vect{x}_{l,m,n}),\\
{\vect g}(\vect x) &=& \sum_{0 \leq l, m, n < M_{B}} {\vect g_{l,m,n}} 
W_{L_{B}}(\vect x - \vect{x}_{l,m,n}),\\
W_{L}(\vect x) &=& \Lambda_{L} (x) ~ \Lambda_{L} (y) ~ \Lambda_{L} (z),
\end{eqnarray}
and
\begin{eqnarray}
\Lambda_{L}(x) &=& \cases{
	1 - l_{L}^{-1} |x|	& if $|x| \leq l_{L};$ \cr
	0			& otherwise,}
\end{eqnarray}
where $l_{L_B}$, $m_p$ and $\vect{x}_{l,m,n}$ are the mesh spacing, the
mass of particles and the position of cell corners, respectively. 
The force on each node is calculated by two-point 
differencing scheme which uses the difference equation for equation 
(\ref{eq:force_diff}):
\begin{eqnarray}
g_{x~l,m,n}= -\frac{1}{2 a l_{L_B}} ( \phi_{l+1, m, n}- \phi_{l-1, m, n}),\\
g_{y~l,m,n}= -\frac{1}{2 a l_{L_B}} ( \phi_{l, m+1, n}- \phi_{l, m-1, n}),
\end{eqnarray}
and
\begin{eqnarray}
g_{z~l,m,n}= -\frac{1}{2 a l_{L_B}} ( \phi_{l, m, n+1}- \phi_{l, m, n-1}).
\end{eqnarray}

%assignment and force interpolation.
The CIC scheme is applied also to the hierarchical meshes. The mass of 
particles is assigned to the corners of all levels of cells 
including the particles, while the force onto the level $L$ particles is 
interpolated from those on the corners of the level $L$ refined cells:
\begin{eqnarray}
\rho_{L}(\vect x_{n}) &=& \sum_{\it p \in P_L} m_p l_{L}^{-3} W_L(\vect x_{p} 
- \vect{x}_{n}),
\end{eqnarray}
and
\begin{eqnarray}
{\vect g_L}(\vect x_{p}) &=& \sum_{n \in N_L} {\vect g_L(\vect x_n)} 
W_L(\vect x_p - \vect{x}_{n}),
\end{eqnarray}
where $P_L$ and $N_L$ are the sets of particles in level $L$ cells and
nodes of level $L$ cells, respectively.
We repeatedly note that the particle's level is defined as the level
of the finest refined cell including the particle. Thus, level $L$
particles can reside in buffer cells whose level is larger than $L$.

% Section 2.5: Poisson solver
\subsection{Poisson solver \label{sec:poisson}}
We adopt the multigrid method (Brandt 1977) to solve the discretized
Poisson equation for the base mesh. Equation (\ref{eq:poisson}) is
discretized as
\begin{eqnarray}
L \phi_{l, m, n} &=& \rho_{l, m, n}, \label{eq:dis poi}
\end{eqnarray}
where $L$ is the discretized Laplacian operator and stands for
\begin{eqnarray*}
L \phi_{l, m, n} &=& \frac{a}{4 \pi G l_{L_B}^2} \times \\
		&&(\phi_{l-1, m, n} + \phi_{l+1, m, n} +\\
		&& \phi_{l, m-1, n} + \phi_{l, m+1, n} +\\
		&& \phi_{l, m, n-1} + \phi_{l, m, n+1} -\\
		&&6~\phi_{l, m, n}),
\end{eqnarray*}
and, the indices $l$, $m$ and $n$ specify the position of nodes on the
base mesh. Since we do not know the exact solution of this equation
initially, we try to estimate it iteratively. The error ($\delta
\psi$) is defined as the difference between the solution ($\phi$) and
the estimate ($\psi$):

\begin{eqnarray}
\delta \psi = \phi - \psi.
\end{eqnarray}
Roughly, $\delta \psi$ can be split into the short and long wavelength
modes. The short wavelength modes converge quickly by the basic
stationary iterative methods, such as the red-black Gauss-Seidel
method which updates $\psi$ by solving equation (\ref{eq:dis poi})
about $\phi_{i, j, k}$ and substituting $\psi$ into $\phi$:

\begin{eqnarray*}
\psi_{l, m, n} = &\frac{1}{6}& (\psi_{l-1, m, n} + \psi_{l+1, m, n} +\\
	&& \psi_{l, m-1, n} + \psi_{l, m+1, n} +\\
	&& \psi_{l, m, n-1} + \psi_{l, m, n+1} -\\
	&& 4 \pi G l_{L_B}^2 a^{-1} \rho_{l, m, n}).
\end{eqnarray*}
This updating of estimate is not carried out all at once, but through two 
steps which do not update the estimate on a node and its six neighboring 
nodes simultaneously. For example, $\psi_{2l, 2m, 2n}$, $\psi_{2l+1, 2m+1, 
2n}$, $\psi_{2l+1, 2m, 2n+1}$ and $\psi_{2l, 2m+1, 2n+1}$ are updated in 
the first step, then $\psi_{2l+1, 2m, 2n}$, $\psi_{2l, 2m+1, 2n}$, 
$\psi_{2l, 2m, 2n+1}$ and $\psi_{2l+1, 2m+1, 2n+1}$ in the second.

On the other hand, the long wavelength modes converge slowly.
This drawback of the iterative methods is overcome in the multigrid method 
as follows: First, the Poisson equation is rewritten as
\begin{eqnarray}
L \phi        &=& L (\psi + \delta \psi) = \rho, \\
L \delta \psi &=& \rho - L \psi  = \delta \rho, \label{eq:mg res}
\end{eqnarray}
where $\delta \rho$ is called the residual. Note that equation (\ref{eq:mg 
res}) relating the error to the residual keeps the same form as the 
original Poisson equation. Next, we prepare meshes for each level from $0$ 
to $L_B$. The residual on the base mesh is projected onto the level $L_B-1$ 
mesh by the fine-to-coarse operator (see Appendix \ref{sec:multigrid}), 
then the error on the base mesh is relaxed using the red-black Gauss-Seidel 
method. In the two-grid method, the error on the level $L_B-1$ mesh is
projected back to the base level and added to the estimate. This
procedure proceeds further in the multigrid method
by estimating the error on the level $L_B-1$ mesh using the level $L_B-2$ 
mesh, and the error on the level $L_B-2$ mesh using the level $L_B-3$ mesh 
and so on.

In the multigrid method described so far, iteration begins from the base 
mesh, but in the full multigrid method, it begins from the level $0$ mesh 
where the solution is given trivially.  Figure \ref{fig:v cycle} 
schematically shows the schedule of a sequence of relaxation,
coarse-to-fine and fine-to-coarse operations. Since the iteration from 
level $L$ to 0 then from 0 to $L$ is V-shaped, this multigrid iteration is 
called a V-cycle. Further description of the multigrid algorithm and the 
terminology used here is found in e.g. \citet{mg_tut} and \citet{recipe}.
\placefigure{fig:v cycle}

%sweep diagram through a step.
In our code, the potential on the hierarchical meshes is solved as
follows: First, the full multigrid method is applied to solve the
Poisson equation on the base mesh. Then the projection of the
potential on the level $L-1$ meshes onto the level $L$ meshes and the
two grid iteration using the level $L-1$ and $L$ meshes are executed
from the level $L_B+1$ meshes to the $L_D$ meshes.

As the particles belonging to the hierarchical cells change the
position, the potential on hierarchical meshes must be modified. After
the level $L$ particles step their position at time $t + (2n + 1)
2^{L_B - L} \Delta t_{L_B}$, provided $0 < (2n + 1) 2^{L_B - L} < 1$,
two grid iteration is successively applied from the level $L-1$ meshes
to the level $L_D$ meshes. Figure \ref{fig:amr v cycle} exhibits this
AMR Poisson solver's schedule for the coarse-to-fine and
fine-to-coarse operations and relaxation in the case of $L_B=2$ and
$L_D=4$.

 The boundary values of the hierarchical meshes are defined on nodes in buffer cells (Fig. \ref{fig:amr boundary}). The potential at the boundary is calculated by the coarse-to-fine operator from the lower level following the full weighting scheme.  Since the potential on the level $L-1$ meshes reflect the density distribution on the level $L$ meshes after the two grid iteration among them, the potential at the boundary is corrected from the initial guess which is projected from the lower level meshes.

We first wrote a code using the nested iteration which applies
the coarse-to-fine projection and the one level red-black Gauss-Seidel
iteration instead of the two grid iteration. But a slight discrepancy
was observed between the correlation functions in the LCDM model
calculated by the nested iteration code with HTS and the code without
HTS, though the difference is negligible between the results
calculated by the two grid iteration code with and without HTS (\S
\ref{sec:lcdm test}). Hence, it is crucial to adopt the two grid or
the multi grid potential solver to implement the HTS technique.

%boundary condition
\placefigure{fig:amr v cycle}
\placefigure{fig:amr boundary}

%##################################################
% Section 3: Tests of the code
%##################################################
\section{Tests of the code \label{sec:test}}
% section 3.1: Force accuracy
\subsection{Force accuracy}
As the first test, we have checked whether the force between particles is 
calculated correctly. First, we placed a truncated singular isothermal
sphere. Following the mesh refinement in accordance with the way
described in \S \ref{sec:hrch mesh}, the force on the particles is calculated.
Then we placed a fiducial particle in a level $L_D$ cell randomly and
recalculate the force on all particles. Subtracting the previously
calculated force from this force, we obtained the gravitational force from
the fiducial particle on the rest of particles. We repeated the above
process 32 times. In figure \ref{fig:force resolution} the 
force from the fiducial particle to the rest of particles is plotted against 
distance between them. Circles, squares, and crosses represent the cases of 
$L_D$= 5, 7, and 9, respectively, while the level of the base mesh is fixed 
as $L_B$=5. In this test, the spacing of the base mesh, the mass of
particles, and the gravitational constant are normalized to
unity. Although the force is softened when the separation is much
smaller than the spacing of the level $L_D$ mesh in each case, the force
coincides with that expected from the Newtonian law at radius
sufficiently larger than this spacing but smaller than the box size.

\placefigure{fig:force resolution}

We also put a homogeneous sphere consists of massless particles added to a
massive particle at the center. Setting $N_{\mbox{\small rfn}}=1$ so
that all particles belong to the level $L_D$ cells, we calculated the
force on the massless particles from the central particle. Since
our code adopts the periodic boundary condition, we compared this
force not with the Newtonian law, but with the sum of the
Newtonian force from periodically placed particles which is calculated
efficiently using the Ewald expansion (Ewald 1921; Sangster \& Dixon
1976). Figure \ref{fig:force error} shows the relative error of the
force calculated by our code in comparison with that calculated by the Ewald
expansion. Circles, squares, and crosses represent the cases of $L_D$=6,
8, and 10, respectively, with $L_B=6$ in common. In any cases, the
error is largest at $r \sim 2 l_{L_D}$ but is kept within 20\%.
\placefigure{fig:force error}

% section 3.2: Single plane wave
\subsection{Single plane wave}
\citet{zld70} formulated the liner approximation for particle trajectories:
\begin{eqnarray}
\vect{x} = \vect{q} + b(t) \vect{p}(\vect{q}),
\end{eqnarray}
where \vect{q} is the Lagrangian position of the particle and $b(t)$
is the growing mode of the linear density perturbation normalized so
that \vect{p}(\vect{q}) gives the initial positional
perturbation. This Zeldovich approximation is not only applicable to
particles in the sub-linear regime, but also gives their exact
trajectories for a system of one-dimensional sheets until any pairs of sheets
crosses each other. Connecting the trajectories before and after sheet
crossings, the trajectories of sheets are calculated with high
accuracy. We test our code by comparing the trajectories calculated by
our code and those calculated by  Yano's one-dimensional code for
which the details are described in Yano \& Gouda (1998).

Initially we have imposed a single sinusoidal plane wave perturbation
to the Einstein-de Sitter (EdS) universe:
\begin{eqnarray}
p(q) &=& A \sin (k q),\\
k    &=& \frac{2 \pi}{l_0},
\end{eqnarray}
and
\begin{eqnarray}
A    &=& 10^{-2} l_{L_B},
\end{eqnarray}
where $l_0$ and $l_{L_B}$ are the size of simulation box and the base mesh 
spacing, respectively. In this test, the level of the base mesh is fixed as 
$L_B=5$.

The following equations for position, velocity, and density hold until the 
first caustic appears:
\begin{eqnarray}
x &=& q + a(t)  A \sin(kq), \\
v &=&\dot{a}(t) A \sin(kq),
\end{eqnarray}
and
\begin{eqnarray}
\rho &=& \bar{\rho} \left(\frac{\partial x}{\partial q}\right)^{-1}\\
     &=& \bar{\rho} (1 + a(t) A k \cos(kq))^{-1},
\end{eqnarray}
where the scale factor $a(t)$ is normalized to unity at the initial
time. When $\rho$ diverges to infinity, the first caustics form. This
occurs at $q = \pi k^{-1} = l_0 / 2$ when $a(t) = a_1 = (Ak)^{-1}$. 
We run the simulations setting $A=0.01$ with sheets consisting of
$L_B^2=25$ particles arranged on the grid. Figure \ref{fig:1stspw} shows
the distribution of sheets in the phase space at $a(t)=a_1$ taken from
the four runs using the one-dimensional code (solid line), the PM code
($L_D=5$, {\it crosses}), our code with HTS ($L_D=7$, {\it circles}),
and our code without HTS ($L_D=7$, {\it squares}). 
Although the cells bordering on the caustics are refined only once, the 
velocity degradation for $L_D=7$ is weaker than that for $L_D=5$.

The difference between the $L_D=$5 and 7 results is more prominent when the 
second caustic appears at $a(t) = a_2 \simeq 2.34 ~ a_1$ (Fig. 
\ref{fig:2ndspw}). Near the center of the spiral pattern in the phase 
space, the $L_D=5$ result is wound weaker than the other two $L_D=7$ 
results. No obvious difference between the $L_D=7$ results with HTS and 
without it shows that our implementation of the HTS is very successful. 
58 base level time steps are used from the beginning of the simulation to
the second caustics generation.
Note that the positions of the first caustics in the three runs are the
same. This is because sheets in low resolution run are pulled toward the
center more weakly than those in high resolution runs, not only when they
fall toward the center, but also when they leave it in the same way.
Thus, the position of caustics does not change much by the resolution of
the code. This is the reason why sheets' position in the phase space is
different among low resolution run and high resolution runs, even though
the position of the caustics are the same. This is true not only for the
plane wave test, but also for the spherical infall test described in the
next.

\placefigure{fig:1stspw}
\placefigure{fig:2ndspw}

% section 3.3: Spherical self-similar infall
\subsection{Spherical self-similar infall}
As the third test, our code is checked against the spherical self-similar 
infall model where the infall of collisionless material onto the collapsed 
density perturbation in the EdS universe is considered. This model is 
solved semi-analytically (Fillmore \& Goldreich 1984; Bertschinger 1985).
As its name indicates, this infall model possesses the symmetry which is 
different from the planar symmetry of the base mesh in our code and the 
single plane wave test.
Moreover, the infall onto the virialized objects is thought to be common in 
the hierarchical clustering. For these reasons, this model is widely used 
as a test for cosmological simulation codes (e.g. Splinter 1996;
Kravtsov et al. 1997; Gelato et al. 1997; Yahagi, Mori, \& Yoshii 1999).

In order to prepare the initial particle distribution for this test 
problem, we place $\Delta N$ particles at the center of particles 
distributed homogeneously and amorphously. Such a distribution is obtained 
as follows \citep{white93}:
First, particles are distributed randomly. Next, the system is evolved in 
the EdS universe reversing the sign of the force to make particles 
repulsive each other, and is expanded million times from the initial time 
until the kinetic energy of the system converges to be negligibly small.

In this test, we set $L_B=7$, $L_D=10$, $\Delta N=64$, and use $2^{3
L_B}=128^3$ particles. We normalize the time by the initial age of the
universe, $\tau = 3 H_i t / 2$, and the radius by the turn-around
radius, $r_{ta}$, given by
\begin{eqnarray}
r_{ta}	&=&	\left(\frac{3\pi}{4}\right)^{-8/9} \delta_i^{1/3} R_i \tau\\
	&=&	\left(\frac{3\pi}{4}\right)^{-8/9}
		\left(\frac{3}{4 \pi} \frac{\Delta N}{N}\right)^{1/3} l_{0} \tau,
\end{eqnarray}
where $N$, $\delta_i$ and $R_i$ are the total number of particles, the 
initial density contrast and its radial extension, respectively. Figure 
\ref{fig:bertschinger test} shows the radial density profile calculated by 
our code ({\it circles}).
Crosses connected by the solid line denote the semi-analytic solution taken 
from Table 4 of \citet{bert85}. Our result shows good coincidence with the 
semi-analytic solution including the position of the first caustic and
the power law asymptote of density profile toward the center whose index is
$-9/4$.

\placefigure{fig:bertschinger test}

% section 3.4: Halos in the CDM universe
\subsection{Halos in the CDM universe \label{sec:lcdm test}}
Finally, we have performed three LCDM simulations, using $64^3$ particles 
and $L_B=6$ in common. We adopt $L_D=6$ for the PM run and $L_D=10$ for 
the AMR+HTS and AMR runs. The difference between the AMR+HTS and AMR 
runs is only that the AMR+HTS run adopts the HTS while the AMR run does 
not. For these simulations, we adopt $\Omega_0 = 0.3, \lambda_0 = 0.7, 
h=0.7$ and $\sigma_8 = 1.0$. The size of simulation box is 70$h^{-1}$ Mpc 
and the mass of particles is $1.5567 \times 10^{11} M_{\odot}$.
The initial data of the simulations are generated by the COSMICS code
provided by Bertschinger \& Bode using the analytic fitting function for
the power spectrum of the CDM universe described in Bardeen et al. (1986).

%density map
Figure \ref{fig:z0map} shows the map of projected density in the 
logarithmic unit at $z=0$. Filtering is not applied to the density maps in any 
panels. The overall mass distributions from the AMR+HTS and AMR runs
agree well with each other, and are more centrally concentrated than
that from the PM run.  This is confirmed by the comparison among the
particle distributions in a slice of 1/16 thickness of the side
(Fig. \ref{fig:z0slice}), and the two-point correlation functions from the
AMR+HTS, AMR, and PM runs (Fig. \ref{fig:crr}). There are many
cosmological $N$-body simulations with parameters similar to our LCDM
simulations. Figure \ref{fig:crr2} shows the correlation function from
our AMR+HTS run ({\it thick dotted line}) overlaid with those from the PM
run in Klypin et al. (1996) ({\it crosses}, hereafter KPM run), AP$^3$M
run in Jenkins et al. (1998) ({\it dashed line}), and ART run in
Col\'{\i}n et al. (1999) ({\it solid line}). Data of overlaid three runs
are taken from figure 4 of Col\'{\i}n et al. (1999). Vertical solid
lines shown with $\epsilon_{\mbox{\tiny AMR+HTS}}$,
$\epsilon_{\mbox{\tiny PM}}$, $\epsilon_{\mbox{\tiny AP}^3\mbox{\tiny
M}}$ represent the force resolution of our AMR+HTS run, KPM run, and the
AP$^3$M run in Jenkins et al. (1998), respectively. Among the three LCDM
simulation runs, the size of the simulation box and the force resolution
of the KPM run is closest to those of our AMR+HTS run. The root mean
square amplitude of the fluctuation at 8 h$^{-1}$Mpc ($\sigma_8$) of our
run is 1.0 while the KPM run adopted $\sigma_8=1.1$. Although the number
of particles used in the AMR+HTS run is $1/64$ of particles in the KPM
run, magnitude of the correlation function at
$r=2 \epsilon_{\mbox{\tiny AMR+HTS}}$ of the AMR+HTS run coincides with
that of the KPM run within 10$\%$.

\placefigure{fig:z0map}
\placefigure{fig:z0slice}
\placefigure{fig:crr}
\placefigure{fig:crr2}

%halo profile of the halo
We have also compared the density profiles of the halos in the AMR+HTS
and AMR runs. Figure \ref{fig:halo_cmp} shows the density profiles of
halos in the AMR+HTS run ({\it circles}) and the AMR run ({\it
squares}). The universal profile proposed by Navarro, Frenk, \& White
(1996),
\begin{eqnarray}
\rho(r) = \frac{\rho_{s}}{(r/r_s)(1+r/r_s)^2}, \label{eq:NFW profile}
\end{eqnarray}
is fitted to the halos in the AMR+HTS run, and the result is denoted by 
the solid line in each panel of Figure \ref{fig:halo_cmp}. This universal 
profile provides a good fit to the dark halos over the entire range of
radius. The profiles of the halos from the AMR+HTS run are in
agreement with those from the AMR run. Slight discrepancies between
them are primarily due to the different positions of satellite halos
in these two runs. This agreement of halo profiles between the AMR+HTS
and AMR runs also indicate that our implementation of the HTS is
successful.
\placefigure{fig:halo_cmp}

%CPU time
Finally we compare the performance of the AMR+HTS and AMR codes
on a PC with an AMD's Athlon 750 MHz processor.  The LCDM
simulations for which we set $L_D-L_B=4$ are performed from
$a$=0.0366 to $a$=1. Thereby the AMR+HTS code uses 133 base-level time steps,
while the AMR code uses 1932 global time steps. The CPU time spent by these
codes is plotted against $a$ in figure 15.  As seen from this
figure, the AMR+HTS code spends less than a half of the CPU
time spent by the AMR code.  Thus the AMR+HTS code saves much
of the CPU time, while keeping the same spatial resolution as
the AMR code.  Higher performance of the AMR+HTS code is achieved
when we set a higher $L_D-L_B$.
\placefigure{fig:cpu}

%##################################################
% Section 4: Summary
%##################################################
\section{Summary \label{sec:summary}}
We have described two techniques of AMR and HTS which are able to increase 
the spatial and time resolutions of the PM method.  The AMR divides and 
removes cells in each time step in compliance with the refinement 
criterion, and enhances the spatial resolution. On the other hand, the HTS 
changes the time step interval depending on the level of particles and 
enhances the time resolution. Three different tests described in \S 
\ref{sec:test} show that our implementation of AMR and HTS is
successful. Compared with the previous codes with AMR, we describe the
HTS technique in detail. It is found that the HTS is an indispensable
technique for the wide dynamic range simulation because importance of
the HTS increases as the finer cells are introduced into the simulation.

%Future works
%Parallelization
We will parallelize our code for massively parallel processors to
realize simulations with more than $10^9$ particles which are needed
to study the links between the formation of galaxies and their
environments (see. Appendix \ref{sec:mass_res}). Sorting the
hierarchical cells by the Morton ordering is a preparatory step for
the parallelization.

%Combining hydrodynamics code
Finally it is worth mentioning that we have already combined our $N$-body 
code with the hydrodynamics code \citep{yy00}. This combined code enables 
us to study the dynamical evolution of both dark matter and baryonic 
components. Including some physical and phenomenological processes such as 
cooling of the gas, star formation, energy feedback from supernovae etc, we 
will trace the formation process of galaxies under the realistic condition 
in the universe.

\acknowledgements
We would like to thank Shu-ichiro Inutsuka for his useful comments on
the Poisson solver of our code and Taihei Yano for providing us the
data for the single plane wave test using his one-dimensional code for
collisionless systems. Some calculations were conducted using the
resources of the the Astronomical Data Analysis Center of the National
Astronomical Observatory, Japan. H.Y. acknowledges the JSPS Research
Fellowships for Young Scientists.  This work has been supported in
part by the Grant-in-Aid for COE research (07CE2002).
\appendix

%##################################################
% Appendix A: #(particles) to resolve
%##################################################
\section{Crude estimation for the number of particles required to
 resolve halos of field galaxies \label{sec:mass_res}}

The relaxation time scale for a halo is given by (see e.g. Binney \&
Tremaine 1987)
\begin{eqnarray}
t_{relax} &\sim& \frac{0.1 N}{\ln N} t_{cross}\\
&\sim& \frac{0.1 N}{\ln N} (G \rho)^{-1/2},
\end{eqnarray}
where $N$ is the number of particles which compose the halo. For a halo 
formed at $z=z_{coll}$ in the Einstein-de Sitter (EdS) universe, we have
\begin{eqnarray}
t_{relax} &\sim& \frac{0.1 N}{\ln N}
	(178 ~ G \rho_0 (1 + z_{coll})^3 )^{-1/2}. \end{eqnarray}
In order to calculate the evolution of the halo correctly, the relaxation 
time scale must exceed the age of the halo at $z=0$, i.e.
\begin{eqnarray}
t_{relax} &\gtrsim& t_{age} \\
	&\gtrsim& \frac{2}{3} H_0^{-1} ( 1 - ( 1 + z_{coll})^{3/2}). 
\end{eqnarray}
The required minimum number of particles, $N_{relax}$, can be estimated
crudely by solving the following non-linear equality:
\begin{eqnarray}
t_{relax}(N_{relax}) = t_{age}.
\end{eqnarray}
Thus the total number of particles, required to keep simulated 
galactic halos unrelaxed, is given by
\begin{eqnarray}
N_{total} &\geq& \frac{M_{total}}{M_{halo}} N_{relax}\\
	  &\simeq& 2.78 ~ \left(\frac{M_{halo}}{10^{11} M_{\odot}}\right)^{-1} 
\left(\frac{L}{1\mbox{Mpc}}\right)^3 \Omega_0 h^2 N_{relax}\\
	  &\sim& 10^9, ~~ (L \sim 32 \mbox{Mpc}, ~ z_{coll} \sim 3, ~ \mbox{EdS})
\end{eqnarray}
where $M_{halo}$ is the minimum mass of halos which do not relax until the 
simulation is completed, and $L$ is the size of simulated region. For the 
reference values of $M_{halo}=10^{11} M_{\odot}, L=32 \mbox{Mpc}, 
z_{coll}=3$, and $(\Omega_0, h)=(1, 0.5)$, we have $N_{total} \gtrsim 
10^{9}$ for the EdS universe.

The above estimate is based on many assumptions. For example, halos are
assumed to have been isolated after they formed, although there are
cluster galaxies which interact with the cluster and other galaxies, and
these interactions can increase the required number of particles.
Hence, this estimation is quite naive and optimistic, but it gives the
lower limit of the number of particles necessary to reproduce
un-relaxed galactic haloes in simulations.
%However, since the PM method with CIC scheme treats particles as cubically extended objects, two body relaxation does not affects in the PM codes as much as collisional codes such as the direct sum method and the tree method.

%##################################################
% Appendix B: pseudo-code
%##################################################
\section{Pseudo-code \label{sec:pseudo-code}}

{\tt
\noindent
$base\_step$ ()\\
\{\\
\hspace*{15mm}int L;\\
\hspace*{15mm}\\
\hspace*{15mm}$time\_steps$();\\
\hspace*{15mm}$velocity\_half\_step$(L$_{\mbox {\tt B}}\leq$L$\leq$L$_{\mbox {\tt D}}$);\\
\hspace*{15mm}if (L$_{\mbox {\tt B}}<$ L$_{\mbox {\tt D}}$)
		$hierarchical\_step$(L$_{\mbox {\tt B}}+1$);\\
\hspace*{15mm}$position\_step$(L$_{\mbox {\tt B}}\leq$L$\leq$L$_{\mbox {\tt D}}$);\\
\hspace*{15mm}$mesh\_modification$(L$_{\mbox {\tt B}}+1\leq$L$\leq$L$_{\mbox {\tt D}}$);\\
\hspace*{15mm}$potential\_solver$(L$_{\mbox {\tt B}}\leq$L$\leq$L$_{\mbox {\tt D}}$);\\
\hspace*{15mm}$velocity\_half\_step$(L$_{\mbox {\tt B}}\leq$L$\leq$L$_{\mbox {\tt D}}$);\\
\}\\
\\
$hierarchical\_step$ (int L$_0$)\\
\{\\
\hspace*{15mm}int L;\\
\hspace*{15mm}\\
\hspace*{15mm}if (L$_0 <$ L$_{\mbox {\tt D}}$)
		$hierarchical\_step$(L$_0+1$);\\
\hspace*{15mm}$position\_step$(L$_0\leq$L$\leq$L$_{\mbox {\tt D}}$);\\
\hspace*{15mm}$mesh\_modification$(L$_0+1\leq$L$\leq$L$_{\mbox {\tt D}}$);\\
\hspace*{15mm}$potential\_solver$(L$_0\leq$L$\leq$L$_{\mbox {\tt D}}$);\\
\hspace*{15mm}$velocity\_full\_step$(L$_0\leq$L$\leq$L$_{\mbox {\tt D}}$);\\
\hspace*{15mm}if (L$_0 < $ L$_{\mbox {\tt D}}$)
		$hierarchical\_step$(L$_0+1$);\\
\}
}

%##########################################################
% Appendix C: Coarse-to-fine and fine-to-coarse operators
%##########################################################
\section{Coarse-to-fine and fine-to-coarse operators\label{sec:multigrid}}

The full-weighting scheme for the fine-to-coarse and the coarse-to-fine 
operators are adopted as
\begin{eqnarray*}
A^{L-1}_{l ,m ,n} = 
	&\frac{1}{8}&A^{L}_{2l, 2m, 2n} + \\
	&\frac{1}{16}&(A^{L}_{2l-1, 2m, 2n} +
			A^{L}_{2l+1, 2m, 2n} +
			A^{L}_{2l, 2m-1, 2n} +
			A^{L}_{2l, 2m+1, 2n} +
			A^{L}_{2l, 2m, 2n-1} +
			A^{L}_{2l, 2m, 2n+1}) +\\
	&\frac{1}{32}&(
			A^{L}_{2l-1, 2m-1, 2n} +
			A^{L}_{2l+1, 2m-1, 2n} +
			A^{L}_{2l-1, 2m+1, 2n} +
			A^{L}_{2l+1, 2m+1, 2n} +\\
		    &&	A^{L}_{2l-1, 2m, 2n-1} +
			A^{L}_{2l+1, 2m, 2n-1} +
			A^{L}_{2l-1, 2m, 2n+1} +
			A^{L}_{2l+1, 2m, 2n+1} +\\
		    &&	A^{L}_{2l, 2m-1, 2n-1} +
			A^{L}_{2l, 2m+1, 2n-1} +
			A^{L}_{2l, 2m-1, 2n+1} +
			A^{L}_{2l, 2m+1, 2n+1}) +\\
	&\frac{1}{64}&(	
			A^{L}_{2l-1, 2m-1, 2n-1} +
			A^{L}_{2l+1, 2m-1, 2n-1} +
			A^{L}_{2l-1, 2m+1, 2n-1} +
			A^{L}_{2l+1, 2m+1, 2n-1} +\\
		    &&	A^{L}_{2l-1, 2m-1, 2n+1} +
			A^{L}_{2l+1, 2m-1, 2n+1} +
			A^{L}_{2l-1, 2m+1, 2n+1} +
			A^{L}_{2l+1, 2m+1, 2n+1}),\\
A^{L+1}_{2l, 2m, 2n} = && A^{L}_{l, m, n},\\
A^{L+1}_{2l+1, 2m, 2n} = &\frac{1}{2}& (
		A^{L}_{l, m, n} + A^{L}_{l+1, m, n}),\\
A^{L+1}_{2l, 2m+1, 2n} = &\frac{1}{2}& (
		A^{L}_{l, m, n} + A^{L}_{l, m+1, n}),\\
A^{L+1}_{2l, 2m, 2n+1} = &\frac{1}{2}& (
		A^{L}_{l, m, n} + A^{L}_{l, m, n+1}),\\
A^{L+1}_{2l+1, 2m+1, 2n} = &\frac{1}{4}& (
		A^{L}_{l, m, n} + A^{L}_{l+1, m, n} + 
		A^{L}_{l, m+1, n} + A^{L}_{l+1, m+1, n}),\\
A^{L+1}_{2l+1, 2m, 2n+1} = &\frac{1}{4}& (
		A^{L}_{l, m, n} + A^{L}_{l+1, m, n} + 
		A^{L}_{l, m, n+1} + A^{L}_{l+1, m, n+1}),\\
A^{L+1}_{2l, 2m+1, 2n+1} = &\frac{1}{4}& (
		A^{L}_{l, m, n} + A^{L}_{l, m+1, n} + 
		A^{L}_{l, m, n+1} + A^{L}_{l, m+1, n+1}),
\end{eqnarray*}
and
\begin{eqnarray*}
A^{L+1}_{2l+1, 2m+1, 2n+1} = &\frac{1}{8}& (
		A^{L}_{l, m, n} + A^{L}_{l+1, m, n} + 
		A^{L}_{l, m+1, n} + A^{L}_{l, m, n+1} +\\
	    &&	A^{L}_{l+1, m+1, n} + A^{L}_{l+1, m, n+1} +
		A^{L}_{l, m+1, n+1} + A^{L}_{l+1, m+1, n+1}).
\end{eqnarray*}
Here, $A^{L}_{i, j, k}$ represents any variables defined on the node which 
is placed at $(x, y, z) = (i~l_{L}, j~l_{L}, k~l_{L})$ on the level $L$ mesh.

\clearpage

\begin{figure}
%\plotone{f4.eps}
\caption{
An example of the hierarchical mesh distribution in the $N$-body code with 
AMR for the case of the LCDM universe described in \S \ref{sec:lcdm test}.
When particles are distributed as shown in the left panel, hierarchical
meshes are placed as shown in the right panel. The shape of the
hierarchical meshes is not geometrically restricted.
\label{fig:eg mesh}}
\end{figure}

\begin{figure}
%\plotone{f5.eps}
\caption{
Schematic illustration of the full multigrid method. The full
multigrid method accelerates the relaxation process by reducing the
long wavelength mode of the residual using the coarser meshes. The
full multigrid iteration begins from the coarsest mesh, i.e., the whole
simulation box. Potential on the level $L_0$ mesh is solved using
level $L$ meshes where $0 \leq L \leq L_0$.
\label{fig:v cycle}}
\end{figure}

\begin{figure}
%\plotone{f6.eps}
\caption{
Schedule of the Poisson solver in our AMR code for the case of $L_B=2$
and $L_D=4$. At time $t$, we execute the full multigrid iteration
first. Then we apply the two grid iteration from $L_B+1$ to
$L_D$. When the level $L$ particles step their position, the two grid
iteration is applied from $L-1$ to $L_D$.
\label{fig:amr v cycle}}
\end{figure}

\begin{figure}
\plotone{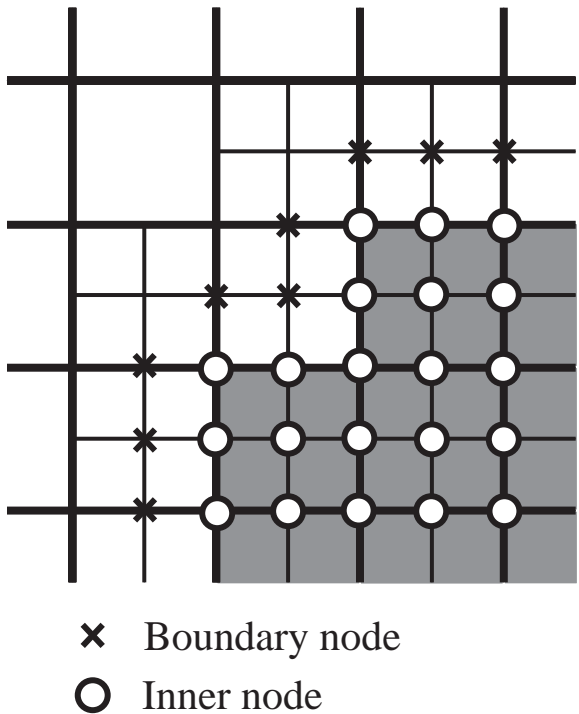}
\caption{
Location of inner nodes ({\it circles}) and outer nodes ({\it crosses}) on 
which the potential and the boundary condition are defined respectively.  
The inner nodes are placed on the corners of refined cells, and the outer 
nodes in buffer cells.
\label{fig:amr boundary}}
\end{figure}

\begin{figure}
%\plotone{f8.eps}
\caption{
Pairwise force in the AMR code. Circles, squares, and crosses are for 
$L_D=$5, 7,  9, respectively, with $L_B=5$ in common.  The size of the
base mesh, the mass of particle, and the gravitational constant are
normalized to unity. The solid line shows the exact Newtonian force of
$g=r^{-2}$. The vertical lines at the bottom indicate the size of the
smallest cell for three cases.
\label{fig:force resolution}}
\end{figure}

\begin{figure}
%\plotone{f9.eps}
\caption{
Error of the pairwise force calculated by the AMR code. Circles,
squares, and crosses are for $L_D=$6, 8,  10, respectively, with
$L_B=6$ in common. Error is defined as the relative error of the
calculated force using the AMR code in comparison with that calculated by the
Ewald expansion. In all cases, the error is maximized at $r \sim 2
l_{L_D}$, but is kept within 20\%.
\label{fig:force error}}
\end{figure}

\begin{figure}
\plotone{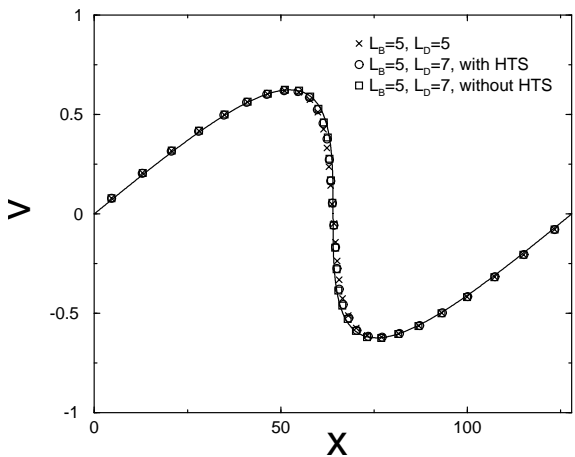}
\caption{
Snapshot of the phase diagram from the single plane wave test at the epoch 
of first caustic generation. The solid line shows the exact solution 
calculated by the one-dimensional code with 1024 sheets using the code
described in Yano \& Gouda (1998).  Crosses, circles, and squares show
the results obtained by the code without AMR, with AMR and HTS, and
with AMR but without HTS, respectively. Because the meshes are refined
only once at this epoch, the difference between the AMR and non-AMR
results is minor. We however note that this difference becomes larger
as time proceeds beyond the first caustic generation (see Fig.
\ref{fig:2ndspw}).
\label{fig:1stspw}}
\end{figure}

\begin{figure}
\plotone{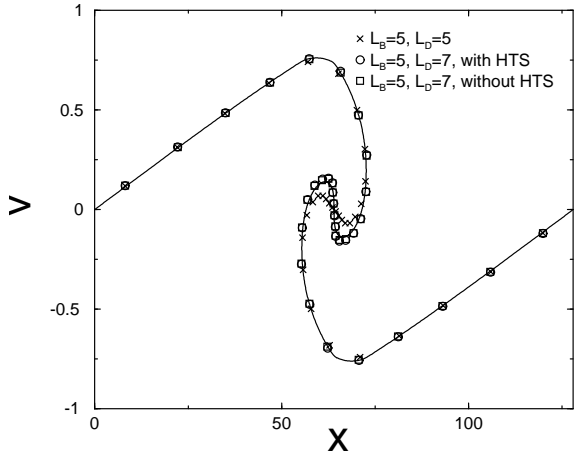}
\caption{
Snapshot of the phase diagram from the single plane wave test at the epoch 
of second caustic generation. The same as figure \ref{fig:1stspw}, but
for $a_2 \simeq 2.34 ~ a_1$ where $a_1$ and $a_2$ are the scale factors
at the epoch of first and second caustics generation, respectively.
Although blunt in the non-AMR run, the second caustics are well captured
in the AMR result, irrespective of whether the HTS is included or not.
\label{fig:2ndspw}}
\end{figure}

\begin{figure}
\plotone{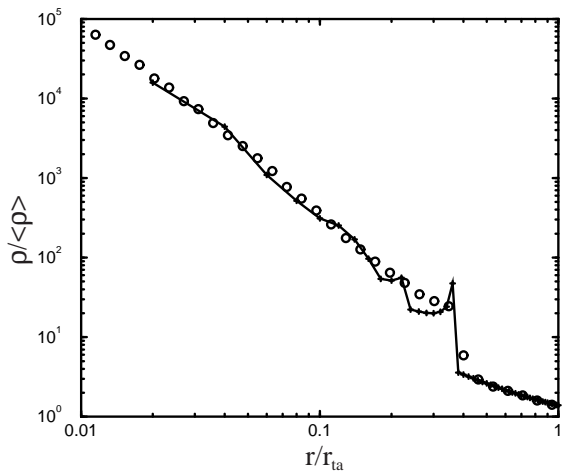}
\caption{
Density profile of the spherical self-similar infall model.  Crosses 
connected by the solid line show the semi-analytic solution provided in 
Table 4 of \citet{bert85}. The circles show the result by our code, 
where radius and density are normalized by the turn around radius and the 
mean density, respectively. Our result shows good agreement with the 
semi-analytic solution including the first caustic.
\label{fig:bertschinger test}}
\end{figure}

\begin{figure}
%\plotone{f13.eps}
\caption{
Map of projected density in the logarithmic unit at $z=0$ for the LCDM 
universe with $\Omega_0 = 0.3, \lambda_0 = 0.7, h=0.7$ and $\sigma_8 = 
1.0$.  The size of simulation box is 70$h^{-1}$ Mpc, and the number of 
particles is $64^3$ and the level of the base mesh, $L_B$, $=6$. Three
panels are taken from (a) AMR run with $L_D=10$, (b) AMR+HTS run with
$L_D=10$, and (c) PM run with $L_D=L_B=6$. The overall mass
distributions of (a) and (b) agree well with each other, and their
halos are bound more tightly than those in (c).
\label{fig:z0map}}
\end{figure}

\begin{figure}
%\plotone{f14.eps}
\caption{
The same as figure \ref{fig:z0map}. Distributions of particles in the
LCDM simulation are shown in this figure. Only particles in a slice
of 1/16 thickness of the simulation box side are shown. This figure
confirms the trends described in the caption of figure \ref{fig:z0map}.
\label{fig:z0slice}}
\end{figure}

\begin{figure}
\plotone{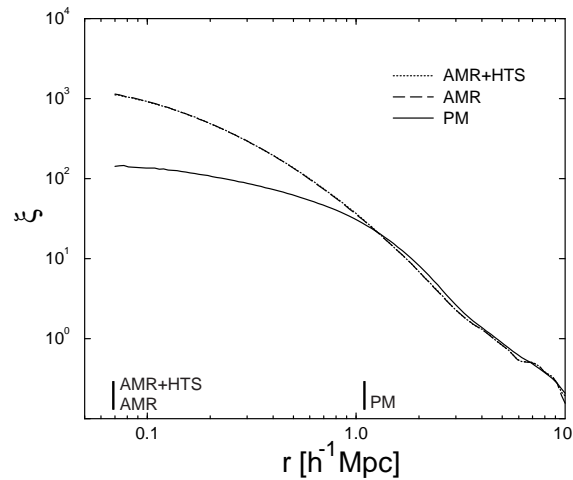}
\caption{
Two-point correlation function at $z=0$ for the LCDM universe with 
$\Omega_0 = 0.3, \lambda_0 = 0.7, h=0.7$ and $\sigma_8 = 1.0$, based on the 
same simulations as in figure \ref{fig:z0map}. Shown are the results from 
the AMR+HTS run ({\it solid line}), the AMR run ({\it dotted line}), and 
the PM run ({\it broken line}).  Note that the AMR+HTS and AMR runs give 
almost identical results, so that the difference is not visible in this 
figure.  The vertical lines indicate the minimum size of the cell, or the 
force resolution of the codes.  Because of the low spatial
resolution the correlation function at small separations in the result
of the PM run is strongly weakened compared with the other two runs. 
\label{fig:crr}}
\end{figure}

\begin{figure}
%\plotone{f16.eps}
\caption{
Comparison of the correlation function of LCDM simulations. Shown are
our AMR+HTS run ({\it thick dotted line}), the PM run in Klypin et
al. (1996) ({\it crosses}), the AP$^3$M run in Jenkins et al. (1998)
({\it dashed line}), and the ART run in Col\'{\i}n et al. (1999) ({\it
solid line}). Vertical solid lines shown with
$\epsilon_{\mbox{\tiny AMR+HTS}}$, $\epsilon_{\mbox{\tiny PM}}$,
$\epsilon_{\mbox{\tiny AP}^3\mbox{\tiny M}}$ represent the force
resolution of our AMR+HTS run, KPM run, and the AP$^3$M run in Jenkins
et al. (1998), respectively.
\label{fig:crr2}}
\end{figure}

\begin{figure}
%\plotone{f17.eps}
\caption{
Density profiles of halos detected at $z=0$ for the LCDM universe with 
$\Omega_0 = 0.3, \lambda_0 = 0.7, h=0.7$ and $\sigma_8 = 1.0$, based on the 
same simulations as in figure \ref{fig:z0map}. Shown are the results from the 
AMR+HTS ({\it circles}) and AMR ({\it squares}) runs.  The solid lines 
show the fits to the halos in the result of the AMR+HTS run using the 
universal density profile (Eq. \ref{eq:NFW profile}).  
\label{fig:halo_cmp}}
\end{figure}

\begin{figure}
\plotone{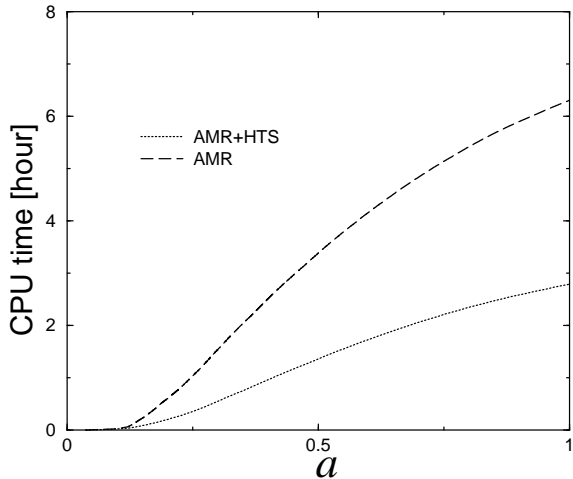}
\caption{
CPU time spent by the AMR+HTS run ({\it dotted line}) and the AMR run
({\it dashed line}) on a PC with an AMD's Athlon 750Hz processor. The
AMR+HTS run spends only a half of the CPU time spent by the AMR run.
\label{fig:cpu}}
\end{figure}
%\figcaption[]{\label{fig:}}
%\figcaption[]{\label{fig:}}
%\figcaption[]{\label{fig:}}
%\figcaption[]{\label{fig:}}
%\figcaption[]{\label{fig:}}
%\figcaption[]{\label{fig:}}
%\figcaption[]{\label{fig:}}
%\figcaption[]{\label{fig:}}
\end{document}